\documentclass[aip,reprint]{revtex4-1}

\usepackage{graphicx}
\usepackage{dcolumn}
\usepackage{bm}
\usepackage[mathlines]{lineno}
\usepackage[utf8]{inputenc}
\usepackage[T1]{fontenc}
\usepackage{mathptmx}
\usepackage{etoolbox}
\usepackage{float}
\usepackage{amsmath}
\usepackage{xcolor}
\draft 
\begin{document}


\title{Laser Writing of Parabolic Micromirrors with a High Numerical Aperture \\for Optical Trapping and Rotation} 

\author{T. Plaskocinski}
\affiliation{SUPA, School of Physics and Astronomy, University of St Andrews, North Haugh, 
St. Andrews, Fife, KY16 9SS, United Kingdom}

\author{Y. Arita}
\affiliation{SUPA, School of Physics and Astronomy, University of St Andrews, North Haugh, 
St. Andrews, Fife, KY16 9SS, United Kingdom}

\author{G. D. Bruce}
\affiliation{SUPA, School of Physics and Astronomy, University of St Andrews, North Haugh, 
St. Andrews, Fife, KY16 9SS, United Kingdom}

\author{S. Persheyev}
\affiliation{SUPA, School of Physics and Astronomy, University of St Andrews, North Haugh, 
St. Andrews, Fife, KY16 9SS, United Kingdom}

\author{K. Dholakia}
\affiliation{SUPA, School of Physics and Astronomy, University of St Andrews, North Haugh,
St. Andrews, Fife, KY16 9SS, United Kingdom}
\affiliation{School of Biological Sciences, University of Adelaide, Adelaide, SA 5005, Australia}
\affiliation{Centre of Light for Life, University of Adelaide, Adelaide, SA 5005, Australia}

\author{A. Di Falco}
\thanks{Authors to whom correspondence should be addressed: adf10@st-andrews.ac.uk, ho35@st-andrews.ac.uk}
\affiliation{SUPA, School of Physics and Astronomy, University of St Andrews, North Haugh, 
St. Andrews, Fife, KY16 9SS, United Kingdom}

\author{H. Ohadi}
\thanks{Authors to whom correspondence should be addressed: adf10@st-andrews.ac.uk, ho35@st-andrews.ac.uk}
\affiliation{SUPA, School of Physics and Astronomy, University of St Andrews, North Haugh, 
St. Andrews, Fife, KY16 9SS, United Kingdom}


\date{\today}

\begin{abstract}
On-chip optical trapping systems allow for high scalability and lower the barrier to access. Systems capable of trapping multiple particles typically come with high cost and complexity. Here we present a technique for making parabolic mirrors with micron-size dimensions and high numerical apertures (NA>1). Over 350 mirrors are made by simple CO\textsubscript{2} laser ablation of glass followed by gold deposition. We fabricate mirrors of arbitrary diameter and depth at a high throughput rate by carefully controlling the ablation parameters. We use the micromirrors for 3-dimensional optical trapping of microbeads in solution, achieving a maximum optical trap stiffness of 52 pN/\textmu m/W. We then further demonstrate the viability of the mirrors as in-situ optical elements through the rotation of a vaterite particle using reflected circularly polarized light. The method used allows for rapid and highly customizable fabrication of dense optical arrays. 
\end{abstract}

\maketitle 

 There has been a great interest\cite{Zhu2019Optofluidics:,Ozcelik2017Optofluidic} in developing optical lab-on-a-chip platforms in recent years as they are portable, robust, 
 and scalable. Integrating multiple optical components has allowed great leaps in analyzing and sensing biological specimens\cite{Padgett2011Holographic,Huang2015SERS-Enabled,Chen2017Biosensors-on-chip:}. 
 One such powerful technique is optical trapping, which allows for the 3D manipulation of objects by using the transfer of 
 momentum of light\cite{Ashkin1970Acceleration}. As a well-established technique, it has found uses in single-molecule force spectroscopy\cite{Campugan2020Optical,Wang1997Stretching}, particle 
 sorting\cite{MacDonald2003Microfluidic,Leake2013Optical}, and sensing in both the far and near field\cite{Bouloumis2020From}. An optical trap is characterized by the trap stiffness $k$, assuming 
 the trapped particle is in a harmonic motion\cite{Neuman2005Optical}. Furthermore, it is possible to transfer the angular momentum of light to 
 anisotropic particles through external or internal birefringence, inducing rotation when trapped by a circularly polarized 
 beam.\cite{Arita2013Laser-induced,Bruce2021Initiating}
 
The system must meet several requirements to allow for the confinement of the subject specimen in an optical potential in the far 
field: the particle must be of a higher refractive index than the surrounding medium and, for optimal performance, should be of a 
size comparable to that of the wavelength of the trapping beam\cite{Metzger2011Observation,Rohrbach2005Stiffness}. In addition, a tightly focused beam is necessary to provide 
a large enough restoring force for trapping. A bulky, high numerical aperture (NA) and high magnification microscope objective is 
required, with multiple drawbacks. For one, the objective size makes the integration difficult, and the high magnification brings 
a small field of view, limiting the number of simultaneously manipulated samples. These and the high cost of high-NA objectives 
have resulted in the poor integration and scalability of the optical trapping platform.

\begin{figure}
\includegraphics[width=8cm]{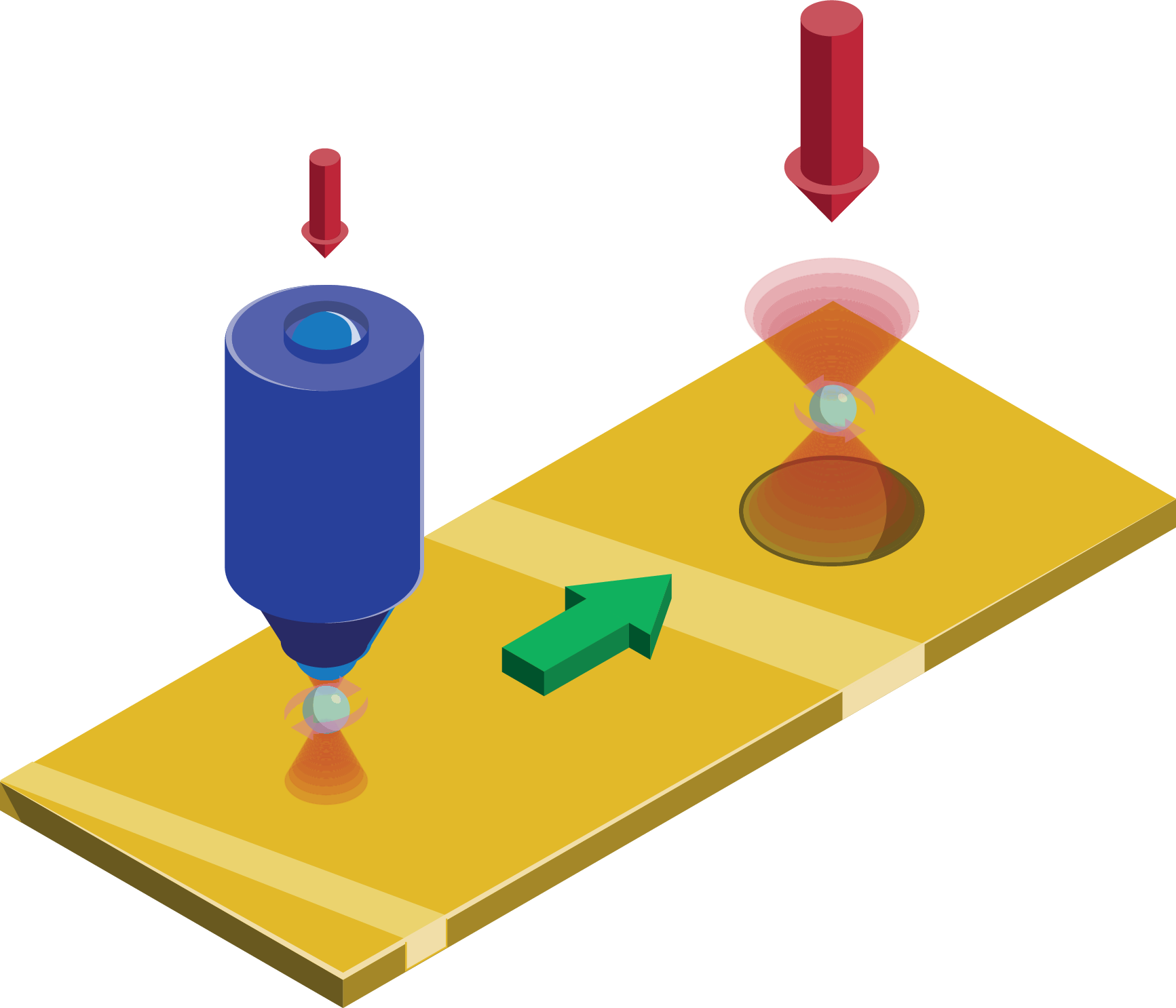}
\caption{\label{fig:fig1} The optical trapping platform schematic. Traditional optical tweezers (left) are used to deliver a bead to micromirror-enabled optical tweezers (right).}
\end{figure}

Multiple schemes have been developed in the last two decades to address these issues to enable on-chip trapping\cite{Paiè2018Particle}. These include 
creating an optical lattice through interference\cite{MacDonald2003Microfluidic}, near and far field trapping enabled by fibers\cite{Lou2019Optical}, as well as working to replace 
the microscope objective with a high numerical aperture but a much smaller optical lens equivalent, capable of similar focusing 
performance. The best performing have been metasurfaces\cite{Xiao2023On-Chip}, placed at the bottom of the microfluidic chamber for trapping in 
reflection\cite{Xiao2023On-Chip,Tkachenko2018Optical}, grafted on top for trapping in transmission\cite{Shen2021On-chip} or placed on tips of optical fibers\cite{Plidschun2021Ultrahigh}. With NA ranging from 0.56 to 
1.2\cite{Markovich2018Optical,Xiao2023On-Chip} , they can reach a performance almost matching that of high NA objectives, all while keeping a small footprint of mm\textsuperscript{2}. 
Another option is the Fresnel diffractive elements\cite{Sun2007Large-scale,Kuo2011Optical,Schonbrun2008Microfabricated}. However, for both systems, the fabrication process is time-consuming 
and expensive. Additionally, the small and delicate optical elements which make up these lenses can be easily damaged. Depending 
on the geometry, they will be wavelength-specific and polarization sensitive. Microlenses and micromirrors\cite{Sow2004Multiple-spot,Merenda2007Multiple}, however, do not 
suffer from these problems. By imprinting pre-made lenses\cite{Zhao2011Microlens-array-enabled,Merenda2009Three-dimensional}, arrays of traps with relatively high NA have been fabricated. 
While convenient, this method is limited by the variety of lenses available to consumers. Chemical treatment can also be combined 
with other techniques to etch smooth mirrors into the substrate\cite{Matsutani2019Microfabrication,Kendall1988,Z_Moktadir_2004,Najer2017}. A method often combined with etching is silica CO\textsubscript{2} laser 
ablation, a well-established glass treatment process whereby a short ($\sim$100 ns) pulse is used to form an approximately 
hemispherical mirror\cite{Ruelle2019Optimized}. Silica is exceptionally absorbent for radiation above 4 \textmu m\cite{Feit2003Mechanisms}, meaning that a focused beam of CO\textsubscript{2} 
laser with a wavelength of 10.6 \textmu m can evaporate the glass directly at the focus, melting the glass around it. The vapor 
pressure from the evaporation then creates a melt front which travels away from the center until the glass re-solidifies, the 
entire process occurring in under a second\cite{Hunger2012Laser,Nowak:06}.

Here, we present a significantly faster, more customizable technique for creating trapping arrays with high numerical aperture. 
Using a continuous wave (CW) CO\textsubscript{2} laser, we write micromirror structures through ablation. First, we expose the glass substrate 
for $\sim$100 ms to a focused TEM\textsubscript{00} mode of the laser, resulting in a nearly parabolic mirror profile\cite{Hunger2010fiber}. The glass is then coated with 
a thin gold layer to achieve a smooth reflective surface. We show that by tuning the ablation parameters, the diameter and depth 
of the micromirrors can be controlled accurately based on the required application. To demonstrate the platform's viability, we 
use a high NA micromirror to trap 5 \textmu m vaterite and 2 \textmu m silica particles suspended in D\textsubscript{2}O. As shown in Fig.~~\ref{fig:fig1}, we 
use a hybrid optical setup that interchangeably allows trapping by a high NA objective or micromirrors. While not necessary to 
enable trapping, it is convenient for this demonstration, as trapping can be performed passively by illuminating using a 
collimated beam. We characterize the optical performance of the mirror, then proceed to use it to reverse the direction of 
rotation of a vaterite particle. Through this, we demonstrate the potential of micromirrors as a simple-to-fabricate and 
versatile lab-on-chip optical platform.

\begin{figure}[ht]
\includegraphics[width=8cm]{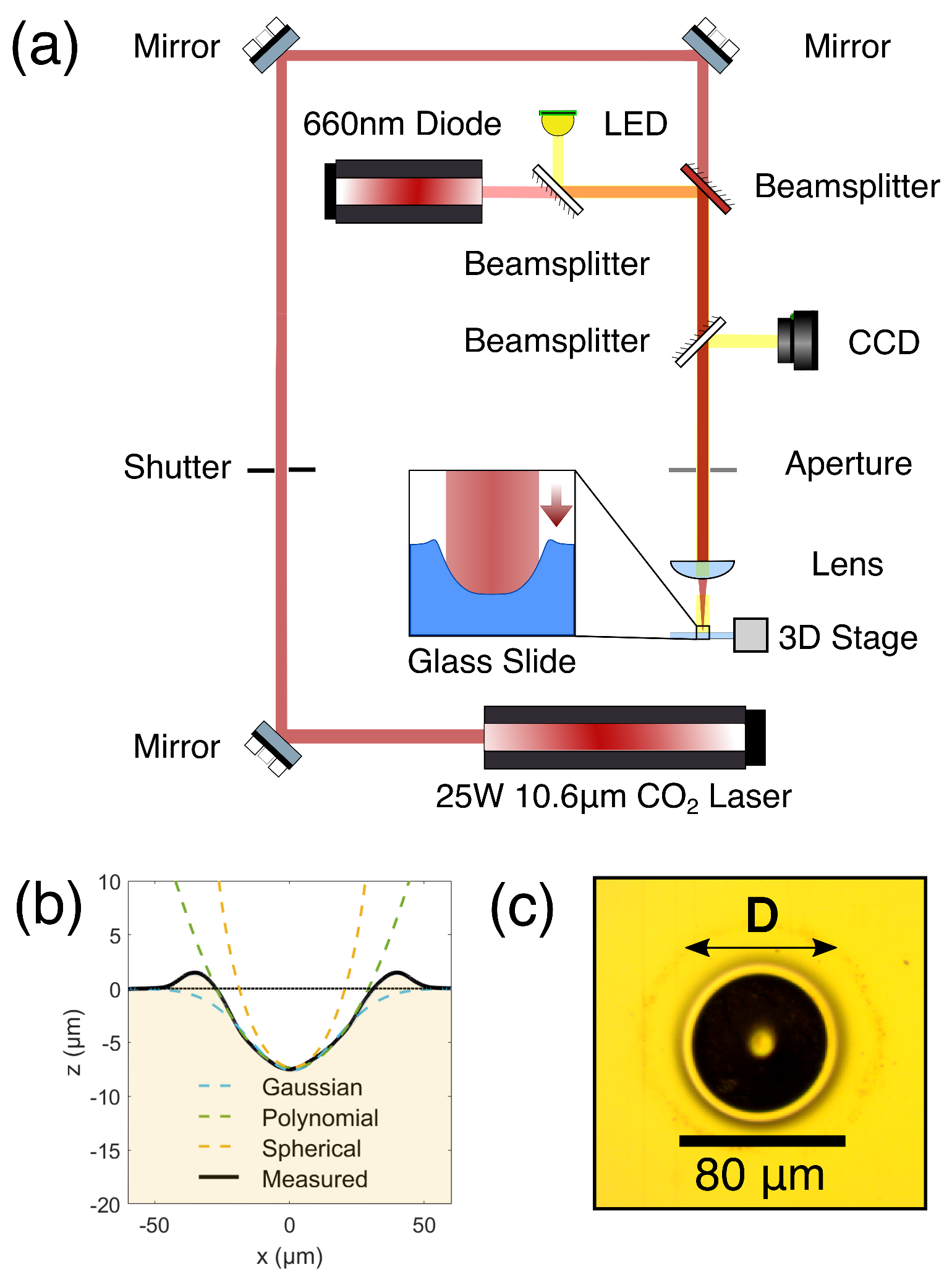}
\caption{\label{fig:fig2} (a) Setup used in micromirror fabrication, with automated shutter, laser, and 3D stage. The lenses and beam splitters for the CO\textsubscript{2} laser were ZnSe-coated to minimize loss. (b) Top-down view of a gold-coated micromirror with diameter D = 80 \textmu m. The bright spot in the center is the reflection of the illuminating lamp. (c) Gaussian, spherical, and polynomial fitting of the mirror profile fabricated with laser power of 1.8 W, single pulse time of $\tau$ = 120 ms, and aperture diameter of 2 mm. We used this mirror for both the trapping and rotation of particles. Refer to Table ~\ref{tab:table1} for all parameters.}
\end{figure}


At the core of the setup for the fabrication of the micromirrors is the CO\textsubscript{2} ablation laser (Synrad 48-KA CW 25 W) (see Fig.~\ref{fig:fig2}(a)). We controlled the duration of the ablation using a mechanical shutter (Thorlabs SC10) and set the average power through 
the duty cycle of the laser ($\sim$8\%, corresponding to $\sim$1.2 W at the sample plane). The glass sample was mounted vertically on a 
motorized 3D stage (Newport, Picomotor actuators). A visible diode laser ($\lambda_D$ = 660 nm) with an adjustable divergence was 
used as a guide to precisely locate the x, y, and z positions of the focused CO\textsubscript{2} laser beam on the sample. For monitoring, we 
used a CMOS camera (Thorlabs DCC1545M) in conjunction with a collimated light-emitting diode. The shutter, CO\textsubscript{2} laser, and stage 
were synchronized and controlled using LabVIEW to rapidly fabricate customized arrays of micromirrors, as shown in S1. The 
automation also allowed for writing continuous channels in glass, as shown in S2. The beam profile of the CO\textsubscript{2} laser was spatially 
filtered using an aperture and finally focused onto the glass sample to a size of $\sim$100 \textmu m using a ZnSe plano-convex lens 
($f$ = 15 mm), which also served as the imaging lens for the camera. After ablation, the sample was transferred to an electron beam 
evaporator (Edwards AUTO 306), where a 2 nm thick adhesion layer of NiCr was deposited, followed by 150 nm of gold.
We characterized the micromirrors using optical microscopy and a surface profiler. We used optical microscopy to extract the 
mirror diameter and depth (Fig.~\ref{fig:fig2}(b)) and a contact mode profilometer (Veeco Dektak 150 with a 12 \textmu m stylus) to accurately 
map the profile (Fig.~\ref{fig:fig2}(c)). The figure shows the micromirror used for particle trapping with depth $z$ = 7.3 \textmu m and 
diameter $D$ = 56.5 \textmu m. The mirror profile resembles the Gaussian intensity profile of the CO\textsubscript{2} beam with deviations due to 
the long pulse durations used, meaning that a parabola more accurately describes the profile\cite{Hunger2012Laser}. The exact scan geometry and 
alternative view of the profile are shown in S3. 
357  micromirrors with diameters of $\sim$40-120 \textmu m were fabricated and characterized. We considered five parameters: the laser 
power, the position of the focus relative to the substrate, the duration and the number of exposures, and the size of the 
aperture (shown in Fig.~\ref{fig:fig2}(a)). We found that the final profile of the micromirror was relatively insensitive to the duration and 
number of exposures, indicating that the shortest exposure created by the shutter ($\sim$50 ms) was longer than the ablation 
timescale. In addition, the diameter of the micromirrors initially displayed a simple linear relationship with the laser power, 
which plateaued as the mirrors' diameter approached that of the focused beam. Finally, the depth of the micromirrors reached an 
upper bound at higher laser power ($\sim$2 W), where the rapid change in temperature gradient resulted in cracks forming at the bottom 
of the mirrors.
Furthermore, the diameter and the depth of the micromirrors had a nearly linear dependence on the aperture diameter. The exact 
position of the focus had the most substantial effect on the micromirror profile. If the mirror's center evaporated too quickly, 
it formed a secondary curvature inside the first, as the melt front would not move fast enough. Hence, the beginning of each 
fabrication included a calibration step where the exact z-plane was chosen based on the mirror shape observed through the CCD 
after ablation. For further information on mirror ablation, see supplementary figures S4 – S7.

\begin{figure}[ht]
\includegraphics[width = 6cm]{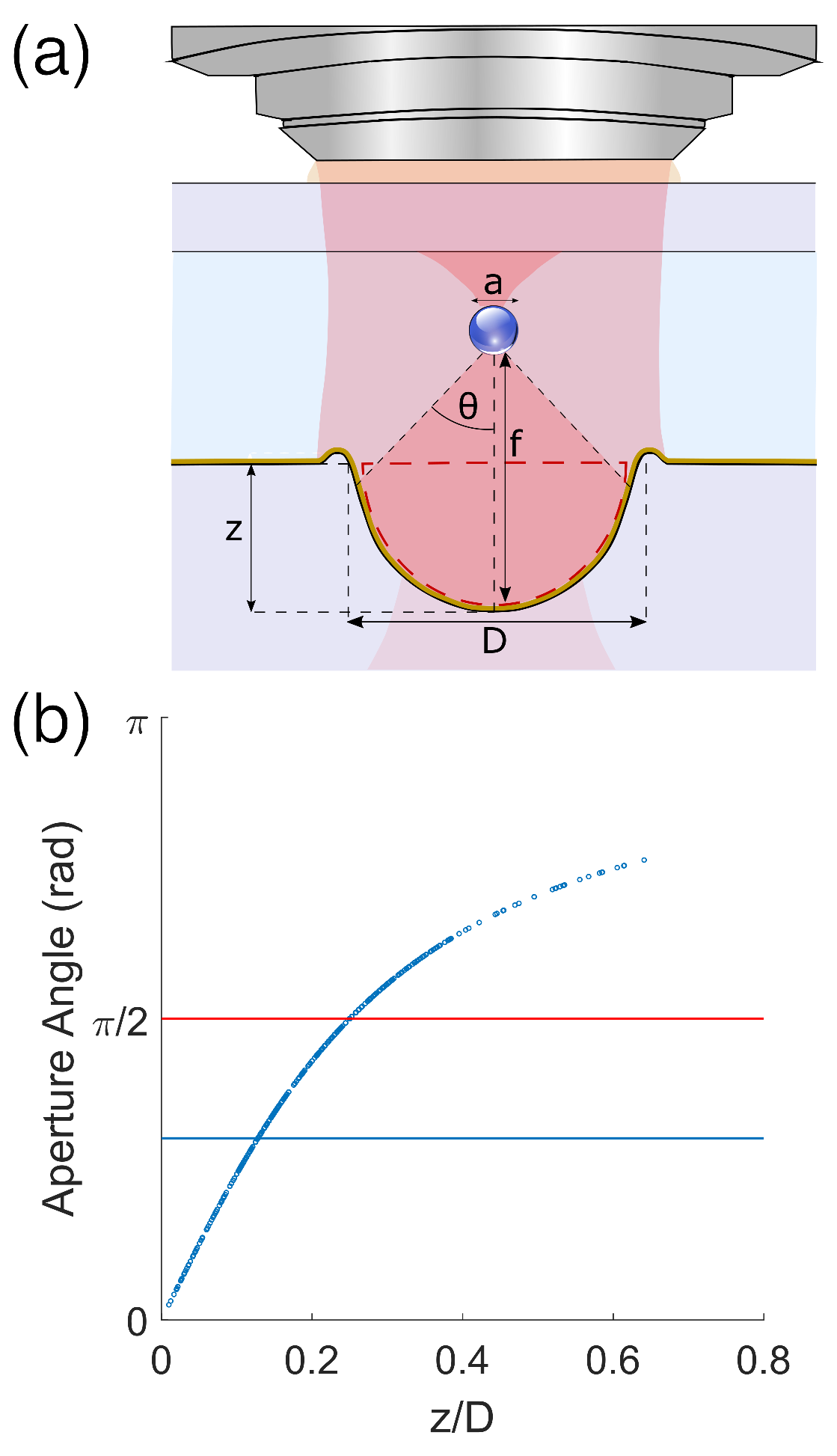}
\caption{\label{fig:fig3}(a) Micromirror trapping scheme, with the depth $z$, diameter $D$, focal length $f$, the radius of curvature $ROC$ and aperture angle $\theta$ as parameters of the mirror, and a is the particle size. (b) Relationship between the aperture angle $\theta$ of the micromirrors and the depth-to-width ratio ($z/D$). The red line is the maximum possible angle before focusing inside the mirror, corresponding to an NA of 1.33 in D\textsubscript{2}O. The blue line indicates the angle of the NA = 1.08 mirror used in the trapping experiment. Each point represents a mirror.}
\end{figure}

To translate the measurable mirror diameter and depth into a usable metric of NA, we used a geometric approximation following 
profile measurements which clearly showed a parabolic profile of the mirror, as seen in Fig.~\ref{fig:fig2}(c). The focal length f of the 
parabola is\cite{Lindlein2007new}:

\begin{equation}
    f = \frac{D^2}{16z},
\end{equation}

where $D$ is the mirror diameter, and $z$ is the depth. We define the aperture angle $\theta$ of a parabolic mirror as:

\begin{equation}
    \frac{z}{f}=\frac{D^2}{16f^2} = \tan^2\left(\frac{\theta}{2}\right).
\end{equation}

We clearly show the relationship between $z/D$ and $\theta$ in Fig.~\ref{fig:fig3}(b), where the parameters of 357  mirrors are plotted 
according to their estimated aperture angle. Of course, this is a simple geometric approximation. Especially for higher ratios of 
$D$ and $z$, the effects of higher fabrication power causing cracks and uneven surfaces must be considered. It is also important 
to note that as $\theta$ increases over $\pi/2$ degrees, the trapping decreases in quality as more of the light contributes to 
the scattering force incident on the particle. For mirrors with $\theta<\pi/2$, a comparison can be made to a lens, using NA as
the figure of merit. Taking the relationship between NA and $\theta$ to be $NA=n_m\sin(\theta)$ where $n_m$ is the refractive
index of the medium (1.33 for D\textsubscript{2}O), we can define an effective NA of the micromirrors defined as:

\begin{equation}
    NA = n_m \sin \left( 2\tan^{-1}\left(\sqrt{\frac{16z^2}{D^2}}\right)\right).
\end{equation}

It is important to note that this comparison with lenses is only valid for mirrors where $\theta\leq\pi/2$, whereas for an 
infinitely extending parabolic mirror, $\theta$ approaches $\pi$.

\begin{figure*}
\includegraphics[width=16cm]{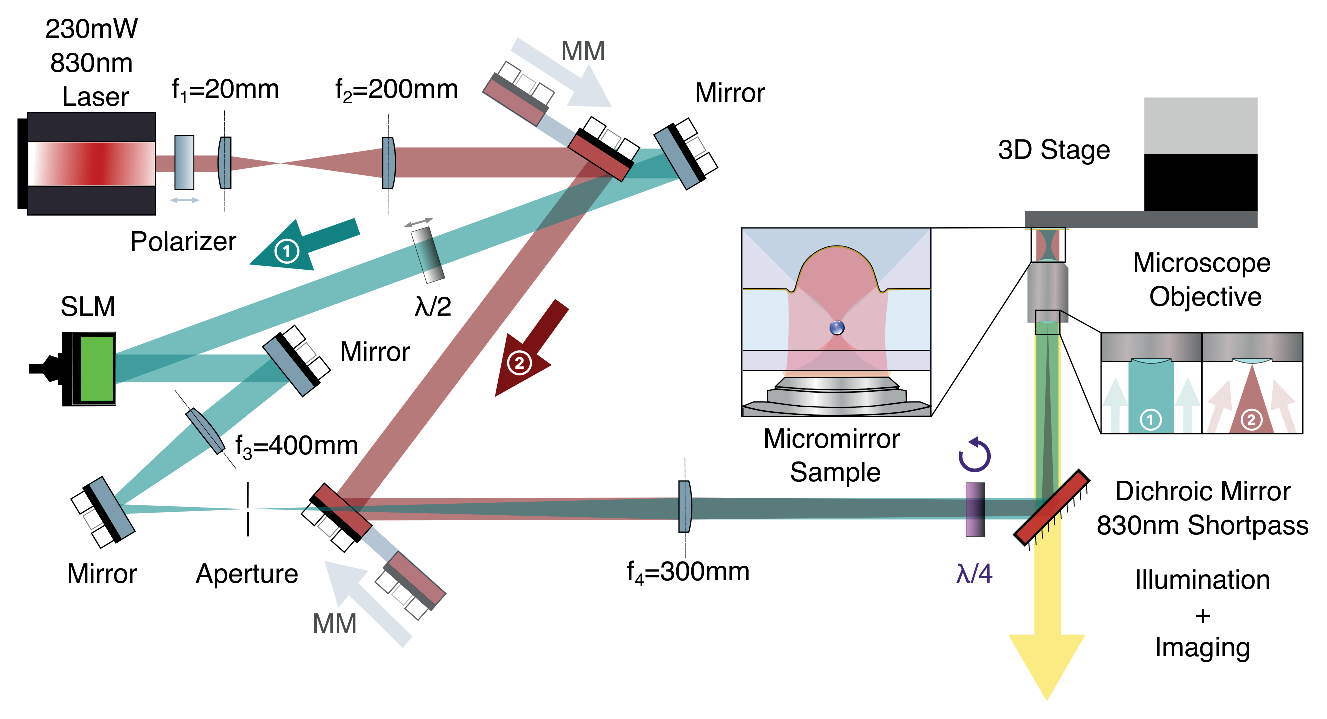}
\caption{\label{fig:fig4}Hybrid optical setup used for trapping particles with the standard optical tweezers using a microscope objective (Path 1, blue) and micromirrors (Path 2, red). The switching between paths is enabled by using movable mirrors (MM)}
\end{figure*}
The hybrid setup used for trapping and rotating the particles is shown in Fig.~\ref{fig:fig4}. The trapping laser (Omicron LuXx) had a total 
power of 230 mW at 830 nm and was linearly polarized (Thorlabs LPNIRB050), after which the beam was expanded by ten times using a 
telescope. The laser beam could then follow one of two paths depending on whether trapping by the objective (Path 1) or 
micromirrors (Path 2) was required. 
Path 1 (marked in blue in Fig.~\ref{fig:fig4}) uses a half-wave plate to match the beam's polarization to that of a spatial light modulator 
(SLM, Meadowlark 1920x1152) for moving the tweezers. The beam was then sent through a 4$f$ system conjugating the plane of the 
SLM and the back focal plane of an inverted water immersion objective (Olympus UPLANSAPO, 60x, NA = 1.2). The use of the SLM was 
not necessary but allowed for the convenient alignment of the particle and mirror illumination when switching paths. For the 
rotation of particles, the beam was circularly polarized using a quarter-wave plate to allow for the transfer of angular momentum 
to birefringent particles.
Path 2 (marked in red) could be switched using a pair of mirrors placed on a motorized linear stage (Thorlabs ELL20). The beam was then focused onto the back of the microscope objective (Fourier plane) 
using a lens ($f_4$ = 300 mm) and the quarter-wave plate for circular polarization. In this way, the incident beam onto the 
micromirrors would be collimated and circularly polarized. The illumination consisted of a fiber-coupled LED, and the 
imaging used a CCD camera (Basler acA640-750um) with an 800 nm short-pass filter in front of it.

The microfluidic chamber containing the mirrors and particles was prepared by placing a 100 \textmu m thick vinyl spacer with a 
hole in the middle to create a well around the mirrors, then depositing the solution of particles, placing a cover slip on top 
and sealing the edges using nail varnish. The schematic is shown in S8. Particle loading on micromirrors consisted of first 
trapping a particle (either silica or vaterite) using the high-NA microscope objective (path 1), then delivering it $\sim$5 \textmu m below the focal point of the micromirrors. Next, path 2 was switched, resulting in a collimated beam illuminating the entire 
micromirror, forming the optical trap (see Supplemental Video SV1). The handover from objective to micromirror would take <1 s. 

\begin{figure}[ht]
\includegraphics[width =8cm]{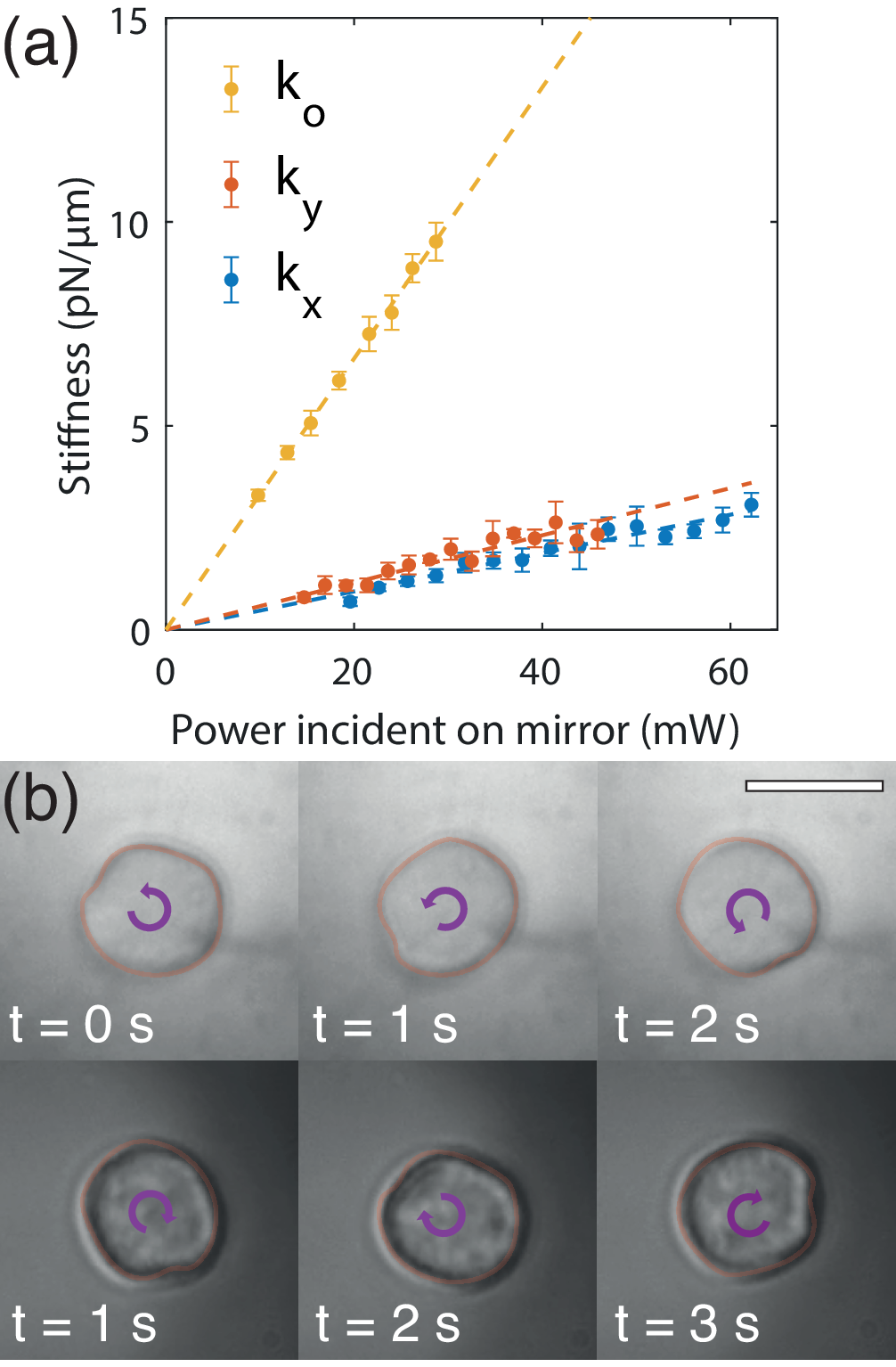}
\caption{\label{fig:fig5} (a) Trap stiffness of the particle in x (blue) and y (orange, and the trap stiffness of the objective with NA = 1.2 (yellow). (b) Rotation of vaterite using traditional optical tweezers (bottom) and the micromirror (top). The scale bar is 5 \textmu m.}
\end{figure}

To quantify the quality of the trapping in the micromirror (see Table ~\ref{tab:table1} for parameters), the mirror's trap stiffness was 
calculated. For all trapping stiffness comparisons, $2.02\pm 0.015$ \textmu m particles (Duke Standards) were used to ensure consistency. First, five videos were taken of the trapped particle at 15 different powers, each video being 10 s long and having a 
framerate of 1000 fps resulting in 50,000 frames containing information about the position variation of the particle for a given 
power. Next, the exact x and y positions of the center of mass of the particles were extracted using a previously demonstrated 
scheme using the shift property of the Fourier transform to symmetrize all images and obtain their relative displacement\cite{Leite2018Three-dimensional}. A 
power spectrum density of the positions was then plotted and fitted with a Lorentzian to acquire a corner frequency $f_c$, which 
was then converted to a trap stiffness through\cite{Neuman2005Optical}:
\begin{equation}
    k_{trap}=2\pi\cdot\underbrace{6\pi\eta a}_{\text{Stokes relation}}\cdot f_c,
\end{equation}

where $\eta$ is the viscosity of D\textsubscript{2}O (1.247 mPa$\cdot$s at room temperature), and $a$ is the particle's radius. Heavy water was used due to a lower absorption coefficient at higher wavelengths\cite{Braun1993}. No proximity correction was necessary as the particle was situated directly in the middle of the chamber. Five stiffness values were obtained for each power, then averaged and plotted against the power measured after the microscope objective (see Fig.~\ref{fig:fig5}(a)). The dashed line is a straight-line fit with no intercept, which yielded a stiffness $k_\perp$ = (52$\pm$1)  pN/\textmu m/W, when correcting for the size of the beam incident on the mirror (see S9). The value is nearly six times lower than that of the objective with NA = 1.2 with a stiffness of $k_o$ = (332$\pm$2 pN/\textmu m/W). An image and diagram of the beam profile can be found in S10, showing a FWHM of $\sim$ 630 nm. Thermal effects are an obvious concern due to the presence of gold: multiple studies have shown that a significant effect from heating is experienced by particles less than a few microns away from the surface, with a focused beam incident on the gold surface\cite{Lu2021Temperature,Xu2018Optical,Wang2022Experimental}. However, in our case, the particle is over 50 \textmu m above the gold surface, the beam is dispersed across an area orders of magnitude larger, and combined with the linear response to changes to the laser power, thermal currents are not expected to have contributed to the trapping dynamics. Also, although the coherence length of our laser would allow for interference, given the relative intensity of the incident collimated beam and the reflective focused beam at the location of the trapping, we exclude the effects of standing waves. Given the wavelength of 830 nm resulting in fringes every 415 nm, and the depth of field of our objective of $\sim$800 nm, we would also have seen the significant displacement of the particle in z, which is absent.

\begin{table}
\caption{\label{tab:table1}Parameters of the micromirror used to trap and rotate particles. In order: diameter, depth, focal length, the ROC, effective numerical aperture, trap stiffness the average of x and y.  }
\begin{ruledtabular}
\begin{tabular}{cccccc}
D (\textmu m) & z (\textmu m) & $f_m$ (\textmu m) & ROC (\textmu m) & NA & $k_\perp$ (pN/\textmu m/W) \\ \hline
56.5         & 7.3          & 54.7                  & 109.4           & 1.08   & 52$\pm$1     

\end{tabular}
\end{ruledtabular}
\end{table}


To further show the potential of our platform as a versatile alternative to a microscope objective, we demonstrated its ability to reverse the preferred direction of rotation of a $\sim$5 \textmu m piece of vaterite. While microvaterite has been shown to reach rotation speeds of up to 5 MHz in vacuum\cite{Arita2013Laser-induced} and up to 400 Hz in water\cite{Bishop2004Optical}, the rate is highly dependent on the power at the sample plane and the particle size (typically in 100s of nm). 
In our demonstration, we place a quarter-wave plate before the objective at 45 degrees with respect to the direction of linear polarization to ensure circular polarization and verify the circularity using a polarimeter (Thorlabs PAX1000IR1) before and after the objective. We first trap the vaterite using objective-enabled optical tweezers and observe the clockwise rotation of the particle, as shown in the bottom panel of Fig.~\ref{fig:fig5}(b) and SV2. We then switch the optical path to trap using the micromirror and observe the particle change in the direction of rotation, as shown in the top panel of Fig.5(b) and SV3. 
To simplify the trapping process, we used relatively large pieces of vaterite ($\sim$5 \textmu m diameter), which were quite asymmetric. These conditions, combined with the low power of our system (<30 mW vs. typical W used\cite{Arita2016Rotational}), means that we observed the particle exhibit a preferred axis of rotation. This is clearly seen in SV2 and SV3 and manifests through a non-uniform rotation rate. We have also considered the role of the angle of reflection on the circular polarization of the reflected beam, as while the effect of reflection from a plane is understood, the case is not as clear for a curved surface such as a parabolic mirror\cite{Mansuripur2011Spin}. While this would be a concern for extreme angles, as can be seen in S3(b) most of the reflections from the mirror happen for $\theta<45^o$, where significant differences between s and p components of the polarization could become apparent.   

In conclusion, we describe the rapid and versatile fabrication process of 357 high-NA parabolic micromirrors with diameters in the range of 80 \textmu m using CW CO\textsubscript{2} laser ablation of silica, followed by gold coating. We show that the micromirrors can be used as a substitute for microscope objectives for optical trapping. We thoroughly analyze the trapping performance at multiple powers, yielding an average stiffness of $k_\perp$ = (52$\pm$1) pN/\textmu m/W for a mirror with an effective NA = 1.08 comparable with values in the literature\cite{Rohrbach2005Stiffness,Yang2021Optical}. Finally, we demonstrate the ability of the mirror to counter-rotate a vaterite particle. The mirror geometry is promising in creating integrated dual-beam trapping\cite{Thalhammer2011Optical}. As closely spaced arrays of arbitrary size can be fabricated, this scheme is a promising technique for on-chip particle trapping and sorting and optical lattices for atoms \cite{Bandi2008Atom,Mu2009Generation,Sondermann2020Prospects}. Similar schemes have also shown great promise regarding the parametric cooling of particles in vacuum\cite{Vovrosh2017Parametric}. With an added laser writing capability, we envisage that guiding particles in arbitrary circuits by 2D trapping in channels would now be feasible in this platform.\\


See Supplementary Information for videos of particle trapping and rotation, fabrication and characterization details.\\

\noindent This work was supported by the UK Engineering and Physical Sciences Research Council (EP/P030017/1 and EP/S014403/1), by the European Research Council (ERC) under the European Union Horizon 2020 Research and Innovation Program (Grant Agreement No. 819346). HO acknowledges support from the Carnegie Trust for Universities of Scotland (Grant No. RIG007685). KD acknowledges support from the Australian Research Council (Grant No. DP220102303).


\noindent The data that support the findings of
this study are openly available in
the Data Set of the University of St. Andrews Research Portal at https://doi.org/10.5194/essd-10-1807-2018. 

\section*{References}
\nocite{*}
\bibliography{Micromirrors_References}

\begin{thebibliography}{54}%
\makeatletter
\providecommand \@ifxundefined [1]{%
 \@ifx{#1\undefined}
}%
\providecommand \@ifnum [1]{%
 \ifnum #1\expandafter \@firstoftwo
 \else \expandafter \@secondoftwo
 \fi
}%
\providecommand \@ifx [1]{%
 \ifx #1\expandafter \@firstoftwo
 \else \expandafter \@secondoftwo
 \fi
}%
\providecommand \natexlab [1]{#1}%
\providecommand \enquote  [1]{``#1''}%
\providecommand \bibnamefont  [1]{#1}%
\providecommand \bibfnamefont [1]{#1}%
\providecommand \citenamefont [1]{#1}%
\providecommand \href@noop [0]{\@secondoftwo}%
\providecommand \href [0]{\begingroup \@sanitize@url \@href}%
\providecommand \@href[1]{\@@startlink{#1}\@@href}%
\providecommand \@@href[1]{\endgroup#1\@@endlink}%
\providecommand \@sanitize@url [0]{\catcode `\\12\catcode `\$12\catcode
  `\&12\catcode `\#12\catcode `\^12\catcode `\_12\catcode `\%12\relax}%
\providecommand \@@startlink[1]{}%
\providecommand \@@endlink[0]{}%
\providecommand \url  [0]{\begingroup\@sanitize@url \@url }%
\providecommand \@url [1]{\endgroup\@href {#1}{\urlprefix }}%
\providecommand \urlprefix  [0]{URL }%
\providecommand \Eprint [0]{\href }%
\providecommand \doibase [0]{http://dx.doi.org/}%
\providecommand \selectlanguage [0]{\@gobble}%
\providecommand \bibinfo  [0]{\@secondoftwo}%
\providecommand \bibfield  [0]{\@secondoftwo}%
\providecommand \translation [1]{[#1]}%
\providecommand \BibitemOpen [0]{}%
\providecommand \bibitemStop [0]{}%
\providecommand \bibitemNoStop [0]{.\EOS\space}%
\providecommand \EOS [0]{\spacefactor3000\relax}%
\providecommand \BibitemShut  [1]{\csname bibitem#1\endcsname}%
\let\auto@bib@innerbib\@empty
\bibitem [{\citenamefont {Zhu}\ \emph {et~al.}(2019)\citenamefont {Zhu},
  \citenamefont {Zhu}, \citenamefont {Zuo}, \citenamefont {Hu}, \citenamefont
  {Shi}, \citenamefont {Liang},\ and\ \citenamefont
  {Yang}}]{Zhu2019Optofluidics:}%
  \BibitemOpen
  \bibfield  {author} {\bibinfo {author} {\bibfnamefont {J.~M.}\ \bibnamefont
  {Zhu}}, \bibinfo {author} {\bibfnamefont {X.~Q.}\ \bibnamefont {Zhu}},
  \bibinfo {author} {\bibfnamefont {Y.~F.}\ \bibnamefont {Zuo}}, \bibinfo
  {author} {\bibfnamefont {X.~J.}\ \bibnamefont {Hu}}, \bibinfo {author}
  {\bibfnamefont {Y.}~\bibnamefont {Shi}}, \bibinfo {author} {\bibfnamefont
  {L.}~\bibnamefont {Liang}}, \ and\ \bibinfo {author} {\bibfnamefont
  {Y.}~\bibnamefont {Yang}},\ }\href@noop {} {\bibfield  {journal} {\bibinfo
  {journal} {Opto-Electronic Advances}\ }\textbf {\bibinfo {volume} {2}},\
  \bibinfo {pages} {1} (\bibinfo {year} {2019})}\BibitemShut {NoStop}%
\bibitem [{\citenamefont {Ozcelik}\ \emph {et~al.}(2017)\citenamefont
  {Ozcelik}, \citenamefont {Cai}, \citenamefont {Leake}, \citenamefont
  {Hawkins},\ and\ \citenamefont {Schmidt}}]{Ozcelik2017Optofluidic}%
  \BibitemOpen
  \bibfield  {author} {\bibinfo {author} {\bibfnamefont {D.}~\bibnamefont
  {Ozcelik}}, \bibinfo {author} {\bibfnamefont {H.}~\bibnamefont {Cai}},
  \bibinfo {author} {\bibfnamefont {K.~D.}\ \bibnamefont {Leake}}, \bibinfo
  {author} {\bibfnamefont {A.~R.}\ \bibnamefont {Hawkins}}, \ and\ \bibinfo
  {author} {\bibfnamefont {H.}~\bibnamefont {Schmidt}},\ }\href {\doibase
  10.1515/nanoph-2016-0156} {\bibfield  {journal} {\bibinfo  {journal}
  {Nanophotonics}\ }\textbf {\bibinfo {volume} {6}},\ \bibinfo {pages} {647}
  (\bibinfo {year} {2017})}\BibitemShut {NoStop}%
\bibitem [{\citenamefont {Padgett}\ and\ \citenamefont
  {Di~Leonardo}(2011)}]{Padgett2011Holographic}%
  \BibitemOpen
  \bibfield  {author} {\bibinfo {author} {\bibfnamefont {M.}~\bibnamefont
  {Padgett}}\ and\ \bibinfo {author} {\bibfnamefont {R.}~\bibnamefont
  {Di~Leonardo}},\ }\href {\doibase 10.1039/c0lc00526f} {\bibfield  {journal}
  {\bibinfo  {journal} {Lab on a Chip}\ }\textbf {\bibinfo {volume} {11}},\
  \bibinfo {pages} {1196} (\bibinfo {year} {2011})}\BibitemShut {NoStop}%
\bibitem [{\citenamefont {Huang}\ \emph {et~al.}(2015)\citenamefont {Huang},
  \citenamefont {Zhang}, \citenamefont {Ding},\ and\ \citenamefont
  {Sun}}]{Huang2015SERS-Enabled}%
  \BibitemOpen
  \bibfield  {author} {\bibinfo {author} {\bibfnamefont {J.~A.}\ \bibnamefont
  {Huang}}, \bibinfo {author} {\bibfnamefont {Y.~L.}\ \bibnamefont {Zhang}},
  \bibinfo {author} {\bibfnamefont {H.}~\bibnamefont {Ding}}, \ and\ \bibinfo
  {author} {\bibfnamefont {H.~B.}\ \bibnamefont {Sun}},\ }\href {\doibase
  10.1002/adom.201400534} {\bibfield  {journal} {\bibinfo  {journal} {Advanced
  Optical Materials}\ }\textbf {\bibinfo {volume} {3}},\ \bibinfo {pages} {618}
  (\bibinfo {year} {2015})}\BibitemShut {NoStop}%
\bibitem [{\citenamefont {Chen}\ and\ \citenamefont
  {Shamsi}(2017)}]{Chen2017Biosensors-on-chip:}%
  \BibitemOpen
  \bibfield  {author} {\bibinfo {author} {\bibfnamefont {S.}~\bibnamefont
  {Chen}}\ and\ \bibinfo {author} {\bibfnamefont {M.~H.}\ \bibnamefont
  {Shamsi}},\ }\href@noop {} {\bibfield  {journal} {\bibinfo  {journal}
  {Journal of Micromechanics and Microengineering}\ }\textbf {\bibinfo {volume}
  {27}} (\bibinfo {year} {2017})}\BibitemShut {NoStop}%
\bibitem [{\citenamefont {Ashkin}(1970)}]{Ashkin1970Acceleration}%
  \BibitemOpen
  \bibfield  {author} {\bibinfo {author} {\bibfnamefont {A.}~\bibnamefont
  {Ashkin}},\ }\href {\doibase 10.1103/PhysRevLett.24.156} {\bibfield
  {journal} {\bibinfo  {journal} {Physical Review Letters}\ }\textbf {\bibinfo
  {volume} {24}},\ \bibinfo {pages} {156} (\bibinfo {year} {1970})}\BibitemShut
  {NoStop}%
\bibitem [{\citenamefont {Campugan}\ \emph {et~al.}(2020)\citenamefont
  {Campugan}, \citenamefont {Dunning},\ and\ \citenamefont
  {Dholakia}}]{Campugan2020Optical}%
  \BibitemOpen
  \bibfield  {author} {\bibinfo {author} {\bibfnamefont {C.~A.}\ \bibnamefont
  {Campugan}}, \bibinfo {author} {\bibfnamefont {K.~R.}\ \bibnamefont
  {Dunning}}, \ and\ \bibinfo {author} {\bibfnamefont {K.}~\bibnamefont
  {Dholakia}},\ }\href {\doibase 10.1080/00107514.2021.1930707} {\bibfield
  {journal} {\bibinfo  {journal} {Contemporary Physics}\ }\textbf {\bibinfo
  {volume} {61}},\ \bibinfo {pages} {277} (\bibinfo {year} {2020})}\BibitemShut
  {NoStop}%
\bibitem [{\citenamefont {Wang}\ \emph {et~al.}(1997)\citenamefont {Wang},
  \citenamefont {Yin}, \citenamefont {Landick}, \citenamefont {Gelles},\ and\
  \citenamefont {Block}}]{Wang1997Stretching}%
  \BibitemOpen
  \bibfield  {author} {\bibinfo {author} {\bibfnamefont {M.~D.}\ \bibnamefont
  {Wang}}, \bibinfo {author} {\bibfnamefont {H.}~\bibnamefont {Yin}}, \bibinfo
  {author} {\bibfnamefont {R.}~\bibnamefont {Landick}}, \bibinfo {author}
  {\bibfnamefont {J.}~\bibnamefont {Gelles}}, \ and\ \bibinfo {author}
  {\bibfnamefont {S.~M.}\ \bibnamefont {Block}},\ }\href {\doibase
  10.1016/S0006-3495(97)78780-0} {\bibfield  {journal} {\bibinfo  {journal}
  {Biophysical Journal}\ }\textbf {\bibinfo {volume} {72}},\ \bibinfo {pages}
  {1335} (\bibinfo {year} {1997})}\BibitemShut {NoStop}%
\bibitem [{\citenamefont {MacDonald}\ \emph {et~al.}(2003)\citenamefont
  {MacDonald}, \citenamefont {Spalding},\ and\ \citenamefont
  {Dholakia}}]{MacDonald2003Microfluidic}%
  \BibitemOpen
  \bibfield  {author} {\bibinfo {author} {\bibfnamefont {M.~P.}\ \bibnamefont
  {MacDonald}}, \bibinfo {author} {\bibfnamefont {G.~C.}\ \bibnamefont
  {Spalding}}, \ and\ \bibinfo {author} {\bibfnamefont {K.}~\bibnamefont
  {Dholakia}},\ }\href {\doibase 10.1038/nature02144} {\bibfield  {journal}
  {\bibinfo  {journal} {Nature}\ }\textbf {\bibinfo {volume} {426}},\ \bibinfo
  {pages} {421} (\bibinfo {year} {2003})}\BibitemShut {NoStop}%
\bibitem [{\citenamefont {Leake}\ \emph {et~al.}(2013)\citenamefont {Leake},
  \citenamefont {Phillips}, \citenamefont {Yuzvinsky}, \citenamefont
  {Hawkins},\ and\ \citenamefont {Schmidt}}]{Leake2013Optical}%
  \BibitemOpen
  \bibfield  {author} {\bibinfo {author} {\bibfnamefont {K.~D.}\ \bibnamefont
  {Leake}}, \bibinfo {author} {\bibfnamefont {B.~S.}\ \bibnamefont {Phillips}},
  \bibinfo {author} {\bibfnamefont {T.~D.}\ \bibnamefont {Yuzvinsky}}, \bibinfo
  {author} {\bibfnamefont {A.~R.}\ \bibnamefont {Hawkins}}, \ and\ \bibinfo
  {author} {\bibfnamefont {H.}~\bibnamefont {Schmidt}},\ }\href {\doibase
  10.1364/oe.21.032605} {\bibfield  {journal} {\bibinfo  {journal} {Optics
  Express}\ }\textbf {\bibinfo {volume} {21}},\ \bibinfo {pages} {32605}
  (\bibinfo {year} {2013})}\BibitemShut {NoStop}%
\bibitem [{\citenamefont {Bouloumis}\ and\ \citenamefont
  {Chormaic}(2020)}]{Bouloumis2020From}%
  \BibitemOpen
  \bibfield  {author} {\bibinfo {author} {\bibfnamefont {T.~D.}\ \bibnamefont
  {Bouloumis}}\ and\ \bibinfo {author} {\bibfnamefont {S.~N.}\ \bibnamefont
  {Chormaic}},\ }\href@noop {} {\bibfield  {journal} {\bibinfo  {journal}
  {Applied Sciences (Switzerland)}\ }\textbf {\bibinfo {volume} {10}} (\bibinfo
  {year} {2020})}\BibitemShut {NoStop}%
\bibitem [{\citenamefont {Neuman}\ and\ \citenamefont
  {Block}(2005)}]{Neuman2005Optical}%
  \BibitemOpen
  \bibfield  {author} {\bibinfo {author} {\bibfnamefont {K.~C.}\ \bibnamefont
  {Neuman}}\ and\ \bibinfo {author} {\bibfnamefont {S.~M.}\ \bibnamefont
  {Block}},\ }\href@noop {} {\bibfield  {journal} {\bibinfo  {journal} {Rev Sci
  Instrum.}\ }\textbf {\bibinfo {volume} {75}} (\bibinfo {year}
  {2005})}\BibitemShut {NoStop}%
\bibitem [{\citenamefont {Arita}\ \emph {et~al.}(2013)\citenamefont {Arita},
  \citenamefont {Mazilu},\ and\ \citenamefont
  {Dholakia}}]{Arita2013Laser-induced}%
  \BibitemOpen
  \bibfield  {author} {\bibinfo {author} {\bibfnamefont {Y.}~\bibnamefont
  {Arita}}, \bibinfo {author} {\bibfnamefont {M.}~\bibnamefont {Mazilu}}, \
  and\ \bibinfo {author} {\bibfnamefont {K.}~\bibnamefont {Dholakia}},\ }\href
  {\doibase 10.1038/ncomms3374} {\bibfield  {journal} {\bibinfo  {journal}
  {Nature Communications}\ }\textbf {\bibinfo {volume} {4}},\ \bibinfo {pages}
  {1} (\bibinfo {year} {2013})}\BibitemShut {NoStop}%
\bibitem [{\citenamefont {Bruce}\ \emph {et~al.}(2021)\citenamefont {Bruce},
  \citenamefont {Rodríguez-Sevilla},\ and\ \citenamefont
  {Dholakia}}]{Bruce2021Initiating}%
  \BibitemOpen
  \bibfield  {author} {\bibinfo {author} {\bibfnamefont {G.~D.}\ \bibnamefont
  {Bruce}}, \bibinfo {author} {\bibfnamefont {P.}~\bibnamefont
  {Rodríguez-Sevilla}}, \ and\ \bibinfo {author} {\bibfnamefont
  {K.}~\bibnamefont {Dholakia}},\ }\href@noop {} {\bibfield  {journal}
  {\bibinfo  {journal} {Advances in Physics: X}\ }\textbf {\bibinfo {volume}
  {6}} (\bibinfo {year} {2021})}\BibitemShut {NoStop}%
\bibitem [{\citenamefont {Metzger}\ \emph {et~al.}(2011)\citenamefont
  {Metzger}, \citenamefont {Mazilu}, \citenamefont {Kelemen}, \citenamefont
  {Ormos},\ and\ \citenamefont {Dholakia}}]{Metzger2011Observation}%
  \BibitemOpen
  \bibfield  {author} {\bibinfo {author} {\bibfnamefont {N.~K.}\ \bibnamefont
  {Metzger}}, \bibinfo {author} {\bibfnamefont {M.}~\bibnamefont {Mazilu}},
  \bibinfo {author} {\bibfnamefont {L.}~\bibnamefont {Kelemen}}, \bibinfo
  {author} {\bibfnamefont {P.}~\bibnamefont {Ormos}}, \ and\ \bibinfo {author}
  {\bibfnamefont {K.}~\bibnamefont {Dholakia}},\ }\href {\doibase
  10.1088/2040-8978/13/4/044018} {\bibfield  {journal} {\bibinfo  {journal}
  {Journal of Optics}\ }\textbf {\bibinfo {volume} {13}},\ \bibinfo {pages}
  {044018} (\bibinfo {year} {2011})}\BibitemShut {NoStop}%
\bibitem [{\citenamefont {Rohrbach}(2005)}]{Rohrbach2005Stiffness}%
  \BibitemOpen
  \bibfield  {author} {\bibinfo {author} {\bibfnamefont {A.}~\bibnamefont
  {Rohrbach}},\ }\href {\doibase 10.1103/PhysRevLett.95.168102} {\bibfield
  {journal} {\bibinfo  {journal} {Physical Review Letters}\ }\textbf {\bibinfo
  {volume} {95}},\ \bibinfo {pages} {168102} (\bibinfo {year}
  {2005})}\BibitemShut {NoStop}%
\bibitem [{\citenamefont {Paiè}\ \emph {et~al.}(2018)\citenamefont {Paiè},
  \citenamefont {Zandrini}, \citenamefont {Vázquez}, \citenamefont
  {Osellame},\ and\ \citenamefont {Bragheri}}]{Paiè2018Particle}%
  \BibitemOpen
  \bibfield  {author} {\bibinfo {author} {\bibfnamefont {P.}~\bibnamefont
  {Paiè}}, \bibinfo {author} {\bibfnamefont {T.}~\bibnamefont {Zandrini}},
  \bibinfo {author} {\bibfnamefont {R.~M.}\ \bibnamefont {Vázquez}}, \bibinfo
  {author} {\bibfnamefont {R.}~\bibnamefont {Osellame}}, \ and\ \bibinfo
  {author} {\bibfnamefont {F.}~\bibnamefont {Bragheri}},\ }\href {\doibase
  10.3390/mi9050200} {\bibfield  {journal} {\bibinfo  {journal}
  {Micromachines}\ }\textbf {\bibinfo {volume} {9}},\ \bibinfo {pages} {1}
  (\bibinfo {year} {2018})}\BibitemShut {NoStop}%
\bibitem [{\citenamefont {Lou}\ \emph {et~al.}(2019)\citenamefont {Lou},
  \citenamefont {Wu},\ and\ \citenamefont {Pang}}]{Lou2019Optical}%
  \BibitemOpen
  \bibfield  {author} {\bibinfo {author} {\bibfnamefont {Y.}~\bibnamefont
  {Lou}}, \bibinfo {author} {\bibfnamefont {D.}~\bibnamefont {Wu}}, \ and\
  \bibinfo {author} {\bibfnamefont {Y.}~\bibnamefont {Pang}},\ }\href {\doibase
  10.1007/s42765-019-00009-8} {\bibfield  {journal} {\bibinfo  {journal}
  {Advanced Fiber Materials}\ }\textbf {\bibinfo {volume} {1}},\ \bibinfo
  {pages} {83} (\bibinfo {year} {2019})}\BibitemShut {NoStop}%
\bibitem [{\citenamefont {Xiao}\ \emph {et~al.}(2023)\citenamefont {Xiao},
  \citenamefont {Plaskocinski}, \citenamefont {Biabanifard}, \citenamefont
  {Persheyev},\ and\ \citenamefont {Di~Falco}}]{Xiao2023On-Chip}%
  \BibitemOpen
  \bibfield  {author} {\bibinfo {author} {\bibfnamefont {J.}~\bibnamefont
  {Xiao}}, \bibinfo {author} {\bibfnamefont {T.}~\bibnamefont {Plaskocinski}},
  \bibinfo {author} {\bibfnamefont {M.}~\bibnamefont {Biabanifard}}, \bibinfo
  {author} {\bibfnamefont {S.}~\bibnamefont {Persheyev}}, \ and\ \bibinfo
  {author} {\bibfnamefont {A.}~\bibnamefont {Di~Falco}},\ }\href@noop {}
  {\bibfield  {journal} {\bibinfo  {journal} {ACS Photonics}\ } (\bibinfo
  {year} {2023})}\BibitemShut {NoStop}%
\bibitem [{\citenamefont {Tkachenko}\ \emph {et~al.}(2018)\citenamefont
  {Tkachenko}, \citenamefont {Stellinga}, \citenamefont {Ruskuc}, \citenamefont
  {Chen}, \citenamefont {Dholakia},\ and\ \citenamefont
  {Krauss}}]{Tkachenko2018Optical}%
  \BibitemOpen
  \bibfield  {author} {\bibinfo {author} {\bibfnamefont {G.}~\bibnamefont
  {Tkachenko}}, \bibinfo {author} {\bibfnamefont {D.}~\bibnamefont
  {Stellinga}}, \bibinfo {author} {\bibfnamefont {A.}~\bibnamefont {Ruskuc}},
  \bibinfo {author} {\bibfnamefont {M.}~\bibnamefont {Chen}}, \bibinfo {author}
  {\bibfnamefont {K.}~\bibnamefont {Dholakia}}, \ and\ \bibinfo {author}
  {\bibfnamefont {T.~F.}\ \bibnamefont {Krauss}},\ }\href {\doibase
  10.1364/ol.43.003224} {\bibfield  {journal} {\bibinfo  {journal} {Optics
  Letters}\ }\textbf {\bibinfo {volume} {43}},\ \bibinfo {pages} {3224}
  (\bibinfo {year} {2018})}\BibitemShut {NoStop}%
\bibitem [{\citenamefont {Shen}\ \emph {et~al.}(2021)\citenamefont {Shen},
  \citenamefont {Duan}, \citenamefont {Ju}, \citenamefont {Xu}, \citenamefont
  {Chen}, \citenamefont {Zhang}, \citenamefont {Ahn}, \citenamefont {Ni},\ and\
  \citenamefont {Li}}]{Shen2021On-chip}%
  \BibitemOpen
  \bibfield  {author} {\bibinfo {author} {\bibfnamefont {K.}~\bibnamefont
  {Shen}}, \bibinfo {author} {\bibfnamefont {Y.}~\bibnamefont {Duan}}, \bibinfo
  {author} {\bibfnamefont {P.}~\bibnamefont {Ju}}, \bibinfo {author}
  {\bibfnamefont {Z.}~\bibnamefont {Xu}}, \bibinfo {author} {\bibfnamefont
  {X.}~\bibnamefont {Chen}}, \bibinfo {author} {\bibfnamefont {L.}~\bibnamefont
  {Zhang}}, \bibinfo {author} {\bibfnamefont {J.}~\bibnamefont {Ahn}}, \bibinfo
  {author} {\bibfnamefont {X.}~\bibnamefont {Ni}}, \ and\ \bibinfo {author}
  {\bibfnamefont {T.}~\bibnamefont {Li}},\ }\href {\doibase
  10.1364/optica.438410} {\bibfield  {journal} {\bibinfo  {journal} {Optica}\
  }\textbf {\bibinfo {volume} {8}},\ \bibinfo {pages} {1359} (\bibinfo {year}
  {2021})}\BibitemShut {NoStop}%
\bibitem [{\citenamefont {Plidschun}\ \emph {et~al.}(2021)\citenamefont
  {Plidschun}, \citenamefont {Ren}, \citenamefont {Kim}, \citenamefont
  {Förster}, \citenamefont {Maier},\ and\ \citenamefont
  {Schmidt}}]{Plidschun2021Ultrahigh}%
  \BibitemOpen
  \bibfield  {author} {\bibinfo {author} {\bibfnamefont {M.}~\bibnamefont
  {Plidschun}}, \bibinfo {author} {\bibfnamefont {H.}~\bibnamefont {Ren}},
  \bibinfo {author} {\bibfnamefont {J.}~\bibnamefont {Kim}}, \bibinfo {author}
  {\bibfnamefont {R.}~\bibnamefont {Förster}}, \bibinfo {author}
  {\bibfnamefont {S.~A.}\ \bibnamefont {Maier}}, \ and\ \bibinfo {author}
  {\bibfnamefont {M.~A.}\ \bibnamefont {Schmidt}},\ }\href@noop {} {\bibfield
  {journal} {\bibinfo  {journal} {Light: Science and Applications}\ }\textbf
  {\bibinfo {volume} {10}} (\bibinfo {year} {2021})}\BibitemShut {NoStop}%
\bibitem [{\citenamefont {Markovich}\ \emph {et~al.}(2018)\citenamefont
  {Markovich}, \citenamefont {Shishkin}, \citenamefont {Hendler},\ and\
  \citenamefont {Ginzburg}}]{Markovich2018Optical}%
  \BibitemOpen
  \bibfield  {author} {\bibinfo {author} {\bibfnamefont {H.}~\bibnamefont
  {Markovich}}, \bibinfo {author} {\bibfnamefont {I.~I.}\ \bibnamefont
  {Shishkin}}, \bibinfo {author} {\bibfnamefont {N.}~\bibnamefont {Hendler}}, \
  and\ \bibinfo {author} {\bibfnamefont {P.}~\bibnamefont {Ginzburg}},\ }\href
  {\doibase 10.1021/acs.nanolett.8b01844} {\bibfield  {journal} {\bibinfo
  {journal} {Nano Letters}\ }\textbf {\bibinfo {volume} {18}},\ \bibinfo
  {pages} {5024} (\bibinfo {year} {2018})}\BibitemShut {NoStop}%
\bibitem [{\citenamefont {Sun}\ \emph {et~al.}(2007)\citenamefont {Sun},
  \citenamefont {Yuan}, \citenamefont {Ong}, \citenamefont {Bu}, \citenamefont
  {Zhu},\ and\ \citenamefont {Liu}}]{Sun2007Large-scale}%
  \BibitemOpen
  \bibfield  {author} {\bibinfo {author} {\bibfnamefont {Y.~Y.}\ \bibnamefont
  {Sun}}, \bibinfo {author} {\bibfnamefont {X.~C.}\ \bibnamefont {Yuan}},
  \bibinfo {author} {\bibfnamefont {L.~S.}\ \bibnamefont {Ong}}, \bibinfo
  {author} {\bibfnamefont {J.}~\bibnamefont {Bu}}, \bibinfo {author}
  {\bibfnamefont {S.~W.}\ \bibnamefont {Zhu}}, \ and\ \bibinfo {author}
  {\bibfnamefont {R.}~\bibnamefont {Liu}},\ }\href {\doibase 10.1063/1.2431768}
  {\bibfield  {journal} {\bibinfo  {journal} {Applied Physics Letters}\
  }\textbf {\bibinfo {volume} {90}},\ \bibinfo {pages} {13} (\bibinfo {year}
  {2007})}\BibitemShut {NoStop}%
\bibitem [{\citenamefont {Kuo}\ and\ \citenamefont
  {Hu}(2011)}]{Kuo2011Optical}%
  \BibitemOpen
  \bibfield  {author} {\bibinfo {author} {\bibfnamefont {J.~N.}\ \bibnamefont
  {Kuo}}\ and\ \bibinfo {author} {\bibfnamefont {H.~Z.}\ \bibnamefont {Hu}},\
  }\href@noop {} {\bibfield  {journal} {\bibinfo  {journal} {Japanese Journal
  of Applied Physics}\ }\textbf {\bibinfo {volume} {50}} (\bibinfo {year}
  {2011})}\BibitemShut {NoStop}%
\bibitem [{\citenamefont {Schonbrun}\ \emph {et~al.}(2008)\citenamefont
  {Schonbrun}, \citenamefont {Rinzler},\ and\ \citenamefont
  {Crozier}}]{Schonbrun2008Microfabricated}%
  \BibitemOpen
  \bibfield  {author} {\bibinfo {author} {\bibfnamefont {E.}~\bibnamefont
  {Schonbrun}}, \bibinfo {author} {\bibfnamefont {C.}~\bibnamefont {Rinzler}},
  \ and\ \bibinfo {author} {\bibfnamefont {K.~B.}\ \bibnamefont {Crozier}},\
  }\href {\doibase 10.1063/1.2837538} {\bibfield  {journal} {\bibinfo
  {journal} {Applied Physics Letters}\ }\textbf {\bibinfo {volume} {92}},\
  \bibinfo {pages} {1} (\bibinfo {year} {2008})}\BibitemShut {NoStop}%
\bibitem [{\citenamefont {Sow}\ \emph {et~al.}(2004)\citenamefont {Sow},
  \citenamefont {Bettiol}, \citenamefont {Lee}, \citenamefont {Cheong},
  \citenamefont {Lim},\ and\ \citenamefont {Watt}}]{Sow2004Multiple-spot}%
  \BibitemOpen
  \bibfield  {author} {\bibinfo {author} {\bibfnamefont {C.~H.}\ \bibnamefont
  {Sow}}, \bibinfo {author} {\bibfnamefont {A.~A.}\ \bibnamefont {Bettiol}},
  \bibinfo {author} {\bibfnamefont {Y.~Y.}\ \bibnamefont {Lee}}, \bibinfo
  {author} {\bibfnamefont {F.~C.}\ \bibnamefont {Cheong}}, \bibinfo {author}
  {\bibfnamefont {C.~T.}\ \bibnamefont {Lim}}, \ and\ \bibinfo {author}
  {\bibfnamefont {F.}~\bibnamefont {Watt}},\ }\href {\doibase
  10.1007/s00340-004-1425-6} {\bibfield  {journal} {\bibinfo  {journal}
  {Applied Physics B: Lasers and Optics}\ }\textbf {\bibinfo {volume} {78}},\
  \bibinfo {pages} {705} (\bibinfo {year} {2004})}\BibitemShut {NoStop}%
\bibitem [{\citenamefont {Merenda}\ \emph {et~al.}(2007)\citenamefont
  {Merenda}, \citenamefont {Rohner},\ and\ \citenamefont
  {Fournier}}]{Merenda2007Multiple}%
  \BibitemOpen
  \bibfield  {author} {\bibinfo {author} {\bibfnamefont {F.}~\bibnamefont
  {Merenda}}, \bibinfo {author} {\bibfnamefont {J.}~\bibnamefont {Rohner}}, \
  and\ \bibinfo {author} {\bibfnamefont {J.-m.}\ \bibnamefont {Fournier}},\
  }\href@noop {} {\bibfield  {journal} {\bibinfo  {journal} {Optics Express}\
  }\textbf {\bibinfo {volume} {15}},\ \bibinfo {pages} {101} (\bibinfo {year}
  {2007})}\BibitemShut {NoStop}%
\bibitem [{\citenamefont {Zhao}\ \emph {et~al.}(2011)\citenamefont {Zhao},
  \citenamefont {Sun}, \citenamefont {Bu}, \citenamefont {Zhu},\ and\
  \citenamefont {Yuan}}]{Zhao2011Microlens-array-enabled}%
  \BibitemOpen
  \bibfield  {author} {\bibinfo {author} {\bibfnamefont {X.}~\bibnamefont
  {Zhao}}, \bibinfo {author} {\bibfnamefont {Y.}~\bibnamefont {Sun}}, \bibinfo
  {author} {\bibfnamefont {J.}~\bibnamefont {Bu}}, \bibinfo {author}
  {\bibfnamefont {S.}~\bibnamefont {Zhu}}, \ and\ \bibinfo {author}
  {\bibfnamefont {X.~C.}\ \bibnamefont {Yuan}},\ }\href {\doibase
  10.1364/AO.50.000318} {\bibfield  {journal} {\bibinfo  {journal} {Applied
  Optics}\ }\textbf {\bibinfo {volume} {50}},\ \bibinfo {pages} {318} (\bibinfo
  {year} {2011})}\BibitemShut {NoStop}%
\bibitem [{\citenamefont {Merenda}\ \emph {et~al.}(2009)\citenamefont
  {Merenda}, \citenamefont {Grossenbacher}, \citenamefont {Jeney},
  \citenamefont {Forró},\ and\ \citenamefont
  {Salathé}}]{Merenda2009Three-dimensional}%
  \BibitemOpen
  \bibfield  {author} {\bibinfo {author} {\bibfnamefont {F.}~\bibnamefont
  {Merenda}}, \bibinfo {author} {\bibfnamefont {M.}~\bibnamefont
  {Grossenbacher}}, \bibinfo {author} {\bibfnamefont {S.}~\bibnamefont
  {Jeney}}, \bibinfo {author} {\bibfnamefont {L.}~\bibnamefont {Forró}}, \
  and\ \bibinfo {author} {\bibfnamefont {R.-p.}\ \bibnamefont {Salathé}},\
  }\href {\doibase 10.1364/ol.34.001063} {\bibfield  {journal} {\bibinfo
  {journal} {Optics letters}\ }\textbf {\bibinfo {volume} {34}},\ \bibinfo
  {pages} {1063} (\bibinfo {year} {2009})}\BibitemShut {NoStop}%
\bibitem [{\citenamefont {Matsutani}\ \emph {et~al.}(2019)\citenamefont
  {Matsutani}, \citenamefont {Sato}, \citenamefont {Hasebe},\ and\
  \citenamefont {Takada}}]{Matsutani2019Microfabrication}%
  \BibitemOpen
  \bibfield  {author} {\bibinfo {author} {\bibfnamefont {A.}~\bibnamefont
  {Matsutani}}, \bibinfo {author} {\bibfnamefont {M.}~\bibnamefont {Sato}},
  \bibinfo {author} {\bibfnamefont {K.}~\bibnamefont {Hasebe}}, \ and\ \bibinfo
  {author} {\bibfnamefont {A.}~\bibnamefont {Takada}},\ }\href {\doibase
  10.18494/SAM.2019.2235} {\bibfield  {journal} {\bibinfo  {journal} {Sensors
  and Materials}\ }\textbf {\bibinfo {volume} {31}},\ \bibinfo {pages} {1325}
  (\bibinfo {year} {2019})}\BibitemShut {NoStop}%
\bibitem [{\citenamefont {Kendall}\ \emph {et~al.}(1988)\citenamefont
  {Kendall}, \citenamefont {de~Guel}, \citenamefont {Guel‐Sandoval},
  \citenamefont {Garcia},\ and\ \citenamefont {Allen}}]{Kendall1988}%
  \BibitemOpen
  \bibfield  {author} {\bibinfo {author} {\bibfnamefont {D.~L.}\ \bibnamefont
  {Kendall}}, \bibinfo {author} {\bibfnamefont {G.~R.}\ \bibnamefont
  {de~Guel}}, \bibinfo {author} {\bibfnamefont {S.}~\bibnamefont
  {Guel‐Sandoval}}, \bibinfo {author} {\bibfnamefont {E.~J.}\ \bibnamefont
  {Garcia}}, \ and\ \bibinfo {author} {\bibfnamefont {T.~A.}\ \bibnamefont
  {Allen}},\ }\href@noop {} {\bibfield  {journal} {\bibinfo  {journal} {Applied
  Physics Letters}\ }\textbf {\bibinfo {volume} {52}},\ \bibinfo {pages} {836}
  (\bibinfo {year} {1988})}\BibitemShut {NoStop}%
\bibitem [{\citenamefont {Moktadir}\ \emph {et~al.}(2004)\citenamefont
  {Moktadir}, \citenamefont {Koukharenka}, \citenamefont {Kraft}, \citenamefont
  {Bagnall}, \citenamefont {Powell}, \citenamefont {Jones},\ and\ \citenamefont
  {Hinds}}]{Z_Moktadir_2004}%
  \BibitemOpen
  \bibfield  {author} {\bibinfo {author} {\bibfnamefont {Z.}~\bibnamefont
  {Moktadir}}, \bibinfo {author} {\bibfnamefont {E.}~\bibnamefont
  {Koukharenka}}, \bibinfo {author} {\bibfnamefont {M.}~\bibnamefont {Kraft}},
  \bibinfo {author} {\bibfnamefont {D.~M.}\ \bibnamefont {Bagnall}}, \bibinfo
  {author} {\bibfnamefont {H.}~\bibnamefont {Powell}}, \bibinfo {author}
  {\bibfnamefont {M.}~\bibnamefont {Jones}}, \ and\ \bibinfo {author}
  {\bibfnamefont {E.~A.}\ \bibnamefont {Hinds}},\ }\href@noop {} {\bibfield
  {journal} {\bibinfo  {journal} {Journal of Micromechanics and
  Microengineering}\ }\textbf {\bibinfo {volume} {14}},\ \bibinfo {pages} {S82}
  (\bibinfo {year} {2004})}\BibitemShut {NoStop}%
\bibitem [{\citenamefont {Najer}\ \emph {et~al.}(2017)\citenamefont {Najer},
  \citenamefont {Renggli}, \citenamefont {Riedel}, \citenamefont
  {Starosielec},\ and\ \citenamefont {Warburton}}]{Najer2017}%
  \BibitemOpen
  \bibfield  {author} {\bibinfo {author} {\bibfnamefont {D.}~\bibnamefont
  {Najer}}, \bibinfo {author} {\bibfnamefont {M.}~\bibnamefont {Renggli}},
  \bibinfo {author} {\bibfnamefont {D.}~\bibnamefont {Riedel}}, \bibinfo
  {author} {\bibfnamefont {S.}~\bibnamefont {Starosielec}}, \ and\ \bibinfo
  {author} {\bibfnamefont {R.~J.}\ \bibnamefont {Warburton}},\ }\href@noop {}
  {\bibfield  {journal} {\bibinfo  {journal} {Applied Physics Letters}\
  }\textbf {\bibinfo {volume} {110}},\ \bibinfo {pages} {011101} (\bibinfo
  {year} {2017})}\BibitemShut {NoStop}%
\bibitem [{\citenamefont {Ruelle}\ \emph {et~al.}(2019)\citenamefont {Ruelle},
  \citenamefont {Poggio},\ and\ \citenamefont
  {Braakman}}]{Ruelle2019Optimized}%
  \BibitemOpen
  \bibfield  {author} {\bibinfo {author} {\bibfnamefont {T.}~\bibnamefont
  {Ruelle}}, \bibinfo {author} {\bibfnamefont {M.}~\bibnamefont {Poggio}}, \
  and\ \bibinfo {author} {\bibfnamefont {F.}~\bibnamefont {Braakman}},\ }\href
  {\doibase 10.1364/ao.58.003784} {\bibfield  {journal} {\bibinfo  {journal}
  {Applied Optics}\ }\textbf {\bibinfo {volume} {58}},\ \bibinfo {pages} {3784}
  (\bibinfo {year} {2019})}\BibitemShut {NoStop}%
\bibitem [{\citenamefont {Feit}\ and\ \citenamefont
  {Rubenchik}(2003)}]{Feit2003Mechanisms}%
  \BibitemOpen
  \bibfield  {author} {\bibinfo {author} {\bibfnamefont {M.~D.}\ \bibnamefont
  {Feit}}\ and\ \bibinfo {author} {\bibfnamefont {A.~M.}\ \bibnamefont
  {Rubenchik}}\ }(\bibinfo {year} {2003})\BibitemShut {NoStop}%
\bibitem [{\citenamefont {Hunger}\ \emph {et~al.}(2012)\citenamefont {Hunger},
  \citenamefont {Deutsch}, \citenamefont {Barbour}, \citenamefont {Warburton},\
  and\ \citenamefont {Reichel}}]{Hunger2012Laser}%
  \BibitemOpen
  \bibfield  {author} {\bibinfo {author} {\bibfnamefont {D.}~\bibnamefont
  {Hunger}}, \bibinfo {author} {\bibfnamefont {C.}~\bibnamefont {Deutsch}},
  \bibinfo {author} {\bibfnamefont {R.~J.}\ \bibnamefont {Barbour}}, \bibinfo
  {author} {\bibfnamefont {R.~J.}\ \bibnamefont {Warburton}}, \ and\ \bibinfo
  {author} {\bibfnamefont {J.}~\bibnamefont {Reichel}},\ }\href@noop {}
  {\bibfield  {journal} {\bibinfo  {journal} {AIP Advances}\ }\textbf {\bibinfo
  {volume} {2}} (\bibinfo {year} {2012})}\BibitemShut {NoStop}%
\bibitem [{\citenamefont {Nowak}\ \emph {et~al.}(2006)\citenamefont {Nowak},
  \citenamefont {Baker},\ and\ \citenamefont {Hall}}]{Nowak:06}%
  \BibitemOpen
  \bibfield  {author} {\bibinfo {author} {\bibfnamefont {K.~M.}\ \bibnamefont
  {Nowak}}, \bibinfo {author} {\bibfnamefont {H.~J.}\ \bibnamefont {Baker}}, \
  and\ \bibinfo {author} {\bibfnamefont {D.~R.}\ \bibnamefont {Hall}},\
  }\href@noop {} {\bibfield  {journal} {\bibinfo  {journal} {Appl. Opt.}\
  }\textbf {\bibinfo {volume} {45}},\ \bibinfo {pages} {162} (\bibinfo {year}
  {2006})}\BibitemShut {NoStop}%
\bibitem [{\citenamefont {Hunger}\ \emph {et~al.}(2010)\citenamefont {Hunger},
  \citenamefont {Steinmetz}, \citenamefont {Colombe}, \citenamefont {Deutsch},
  \citenamefont {Hänsch},\ and\ \citenamefont {Reichel}}]{Hunger2010fiber}%
  \BibitemOpen
  \bibfield  {author} {\bibinfo {author} {\bibfnamefont {D.}~\bibnamefont
  {Hunger}}, \bibinfo {author} {\bibfnamefont {T.}~\bibnamefont {Steinmetz}},
  \bibinfo {author} {\bibfnamefont {Y.}~\bibnamefont {Colombe}}, \bibinfo
  {author} {\bibfnamefont {C.}~\bibnamefont {Deutsch}}, \bibinfo {author}
  {\bibfnamefont {T.~W.}\ \bibnamefont {Hänsch}}, \ and\ \bibinfo {author}
  {\bibfnamefont {J.}~\bibnamefont {Reichel}},\ }\href@noop {} {\bibfield
  {journal} {\bibinfo  {journal} {New Journal of Physics}\ }\textbf {\bibinfo
  {volume} {12}} (\bibinfo {year} {2010})}\BibitemShut {NoStop}%
\bibitem [{\citenamefont {Lindlein}\ \emph {et~al.}(2007)\citenamefont
  {Lindlein}, \citenamefont {Maiwald}, \citenamefont {Konermann}, \citenamefont
  {Sondermann}, \citenamefont {Peschel},\ and\ \citenamefont
  {Leuchs}}]{Lindlein2007new}%
  \BibitemOpen
  \bibfield  {author} {\bibinfo {author} {\bibfnamefont {N.}~\bibnamefont
  {Lindlein}}, \bibinfo {author} {\bibfnamefont {R.}~\bibnamefont {Maiwald}},
  \bibinfo {author} {\bibfnamefont {H.}~\bibnamefont {Konermann}}, \bibinfo
  {author} {\bibfnamefont {M.}~\bibnamefont {Sondermann}}, \bibinfo {author}
  {\bibfnamefont {U.}~\bibnamefont {Peschel}}, \ and\ \bibinfo {author}
  {\bibfnamefont {G.}~\bibnamefont {Leuchs}},\ }\href {\doibase
  10.1134/S1054660X07070055} {\bibfield  {journal} {\bibinfo  {journal} {Laser
  Physics}\ }\textbf {\bibinfo {volume} {17}},\ \bibinfo {pages} {927}
  (\bibinfo {year} {2007})}\BibitemShut {NoStop}%
\bibitem [{\citenamefont {Leite}\ \emph {et~al.}(2018)\citenamefont {Leite},
  \citenamefont {Turtaev}, \citenamefont {Jiang}, \citenamefont {Šiler},
  \citenamefont {Cuschieri}, \citenamefont {Russell},\ and\ \citenamefont
  {Čižmár}}]{Leite2018Three-dimensional}%
  \BibitemOpen
  \bibfield  {author} {\bibinfo {author} {\bibfnamefont {I.~T.}\ \bibnamefont
  {Leite}}, \bibinfo {author} {\bibfnamefont {S.}~\bibnamefont {Turtaev}},
  \bibinfo {author} {\bibfnamefont {X.}~\bibnamefont {Jiang}}, \bibinfo
  {author} {\bibfnamefont {M.}~\bibnamefont {Šiler}}, \bibinfo {author}
  {\bibfnamefont {A.}~\bibnamefont {Cuschieri}}, \bibinfo {author}
  {\bibfnamefont {P.~S.~J.}\ \bibnamefont {Russell}}, \ and\ \bibinfo {author}
  {\bibfnamefont {T.}~\bibnamefont {Čižmár}},\ }\href {\doibase
  10.1038/s41566-017-0053-8} {\bibfield  {journal} {\bibinfo  {journal} {Nature
  Photonics}\ }\textbf {\bibinfo {volume} {12}},\ \bibinfo {pages} {33}
  (\bibinfo {year} {2018})}\BibitemShut {NoStop}%
\bibitem [{\citenamefont {Braun}\ and\ \citenamefont
  {Smirnov}(1993)}]{Braun1993}%
  \BibitemOpen
  \bibfield  {author} {\bibinfo {author} {\bibfnamefont {C.~L.}\ \bibnamefont
  {Braun}}\ and\ \bibinfo {author} {\bibfnamefont {S.~N.}\ \bibnamefont
  {Smirnov}},\ }\href@noop {} {\bibfield  {journal} {\bibinfo  {journal}
  {Journal of Chemical Education}\ }\textbf {\bibinfo {volume} {70}},\ \bibinfo
  {pages} {612} (\bibinfo {year} {1993})}\BibitemShut {NoStop}%
\bibitem [{\citenamefont {Lu}\ \emph {et~al.}(2021)\citenamefont {Lu},
  \citenamefont {Gámez},\ and\ \citenamefont
  {Haro-González}}]{Lu2021Temperature}%
  \BibitemOpen
  \bibfield  {author} {\bibinfo {author} {\bibfnamefont {D.}~\bibnamefont
  {Lu}}, \bibinfo {author} {\bibfnamefont {F.}~\bibnamefont {Gámez}}, \ and\
  \bibinfo {author} {\bibfnamefont {P.}~\bibnamefont {Haro-González}},\ }\href
  {\doibase 10.3390/mi12080954} {\bibfield  {journal} {\bibinfo  {journal}
  {Micromachines}\ }\textbf {\bibinfo {volume} {12}},\ \bibinfo {pages} {1}
  (\bibinfo {year} {2021})}\BibitemShut {NoStop}%
\bibitem [{\citenamefont {Xu}\ \emph {et~al.}(2018)\citenamefont {Xu},
  \citenamefont {Song},\ and\ \citenamefont {Crozier}}]{Xu2018Optical}%
  \BibitemOpen
  \bibfield  {author} {\bibinfo {author} {\bibfnamefont {Z.}~\bibnamefont
  {Xu}}, \bibinfo {author} {\bibfnamefont {W.}~\bibnamefont {Song}}, \ and\
  \bibinfo {author} {\bibfnamefont {K.~B.}\ \bibnamefont {Crozier}},\ }\href
  {\doibase 10.1021/acsphotonics.8b01250} {\bibfield  {journal} {\bibinfo
  {journal} {ACS Photonics}\ }\textbf {\bibinfo {volume} {5}},\ \bibinfo
  {pages} {4993} (\bibinfo {year} {2018})}\BibitemShut {NoStop}%
\bibitem [{\citenamefont {Wang}\ \emph {et~al.}(2022)\citenamefont {Wang},
  \citenamefont {Bai}, \citenamefont {Zhang}, \citenamefont {Li}, \citenamefont
  {Ji},\ and\ \citenamefont {Zhong}}]{Wang2022Experimental}%
  \BibitemOpen
  \bibfield  {author} {\bibinfo {author} {\bibfnamefont {H.~D.}\ \bibnamefont
  {Wang}}, \bibinfo {author} {\bibfnamefont {W.}~\bibnamefont {Bai}}, \bibinfo
  {author} {\bibfnamefont {B.}~\bibnamefont {Zhang}}, \bibinfo {author}
  {\bibfnamefont {B.~W.}\ \bibnamefont {Li}}, \bibinfo {author} {\bibfnamefont
  {F.}~\bibnamefont {Ji}}, \ and\ \bibinfo {author} {\bibfnamefont {M.~C.}\
  \bibnamefont {Zhong}},\ }\href@noop {} {\bibfield  {journal} {\bibinfo
  {journal} {Photonics}\ }\textbf {\bibinfo {volume} {9}} (\bibinfo {year}
  {2022})}\BibitemShut {NoStop}%
\bibitem [{\citenamefont {Bishop}\ \emph {et~al.}(2004)\citenamefont {Bishop},
  \citenamefont {Nieminen}, \citenamefont {Heckenberg},\ and\ \citenamefont
  {Rubinsztein-Dunlop}}]{Bishop2004Optical}%
  \BibitemOpen
  \bibfield  {author} {\bibinfo {author} {\bibfnamefont {A.~I.}\ \bibnamefont
  {Bishop}}, \bibinfo {author} {\bibfnamefont {T.~A.}\ \bibnamefont
  {Nieminen}}, \bibinfo {author} {\bibfnamefont {N.~R.}\ \bibnamefont
  {Heckenberg}}, \ and\ \bibinfo {author} {\bibfnamefont {H.}~\bibnamefont
  {Rubinsztein-Dunlop}},\ }\href {\doibase 10.1103/PhysRevLett.92.198104}
  {\bibfield  {journal} {\bibinfo  {journal} {Physical Review Letters}\
  }\textbf {\bibinfo {volume} {92}},\ \bibinfo {pages} {14} (\bibinfo {year}
  {2004})}\BibitemShut {NoStop}%
\bibitem [{\citenamefont {Arita}\ \emph {et~al.}(2016)\citenamefont {Arita},
  \citenamefont {Richards}, \citenamefont {Mazilu}, \citenamefont {Spalding},
  \citenamefont {Skelton~Spesyvtseva}, \citenamefont {Craig},\ and\
  \citenamefont {Dholakia}}]{Arita2016Rotational}%
  \BibitemOpen
  \bibfield  {author} {\bibinfo {author} {\bibfnamefont {Y.}~\bibnamefont
  {Arita}}, \bibinfo {author} {\bibfnamefont {J.~M.}\ \bibnamefont {Richards}},
  \bibinfo {author} {\bibfnamefont {M.}~\bibnamefont {Mazilu}}, \bibinfo
  {author} {\bibfnamefont {G.~C.}\ \bibnamefont {Spalding}}, \bibinfo {author}
  {\bibfnamefont {S.~E.}\ \bibnamefont {Skelton~Spesyvtseva}}, \bibinfo
  {author} {\bibfnamefont {D.}~\bibnamefont {Craig}}, \ and\ \bibinfo {author}
  {\bibfnamefont {K.}~\bibnamefont {Dholakia}},\ }\href@noop {} {\bibfield
  {journal} {\bibinfo  {journal} {ACS Nano}\ }\textbf {\bibinfo {volume}
  {10}},\ \bibinfo {pages} {11505} (\bibinfo {year} {2016})}\BibitemShut
  {NoStop}%
\bibitem [{\citenamefont {Mansuripur}\ \emph {et~al.}(2011)\citenamefont
  {Mansuripur}, \citenamefont {Zakharian},\ and\ \citenamefont
  {Wright}}]{Mansuripur2011Spin}%
  \BibitemOpen
  \bibfield  {author} {\bibinfo {author} {\bibfnamefont {M.}~\bibnamefont
  {Mansuripur}}, \bibinfo {author} {\bibfnamefont {A.~R.}\ \bibnamefont
  {Zakharian}}, \ and\ \bibinfo {author} {\bibfnamefont {E.~M.}\ \bibnamefont
  {Wright}},\ }\href {\doibase 10.1103/PhysRevA.84.033813} {\bibfield
  {journal} {\bibinfo  {journal} {Physical Review A - Atomic, Molecular, and
  Optical Physics}\ }\textbf {\bibinfo {volume} {84}},\ \bibinfo {pages} {1}
  (\bibinfo {year} {2011})}\BibitemShut {NoStop}%
\bibitem [{\citenamefont {Yang}\ \emph {et~al.}(2021)\citenamefont {Yang},
  \citenamefont {Ren}, \citenamefont {Chen}, \citenamefont {Arita},\ and\
  \citenamefont {Rosales-Guzmán}}]{Yang2021Optical}%
  \BibitemOpen
  \bibfield  {author} {\bibinfo {author} {\bibfnamefont {Y.}~\bibnamefont
  {Yang}}, \bibinfo {author} {\bibfnamefont {Y.~X.}\ \bibnamefont {Ren}},
  \bibinfo {author} {\bibfnamefont {M.}~\bibnamefont {Chen}}, \bibinfo {author}
  {\bibfnamefont {Y.}~\bibnamefont {Arita}}, \ and\ \bibinfo {author}
  {\bibfnamefont {C.}~\bibnamefont {Rosales-Guzmán}},\ }\href {\doibase
  10.1117/1.AP.3.3.034001} {\bibfield  {journal} {\bibinfo  {journal} {Advanced
  Photonics}\ }\textbf {\bibinfo {volume} {3}},\ \bibinfo {pages} {1} (\bibinfo
  {year} {2021})}\BibitemShut {NoStop}%
\bibitem [{\citenamefont {Thalhammer}\ \emph {et~al.}(2011)\citenamefont
  {Thalhammer}, \citenamefont {Steiger},\ and\ \citenamefont
  {Bernet}}]{Thalhammer2011Optical}%
  \BibitemOpen
  \bibfield  {author} {\bibinfo {author} {\bibfnamefont {G.}~\bibnamefont
  {Thalhammer}}, \bibinfo {author} {\bibfnamefont {R.}~\bibnamefont {Steiger}},
  \ and\ \bibinfo {author} {\bibfnamefont {S.}~\bibnamefont {Bernet}},\
  }\href@noop {} {\bibfield  {journal} {\bibinfo  {journal} {Journal of
  Optics}\ }\textbf {\bibinfo {volume} {14}} (\bibinfo {year}
  {2011})}\BibitemShut {NoStop}%
\bibitem [{\citenamefont {Bandi}\ \emph {et~al.}(2008)\citenamefont {Bandi},
  \citenamefont {Minogin},\ and\ \citenamefont {Chormaic}}]{Bandi2008Atom}%
  \BibitemOpen
  \bibfield  {author} {\bibinfo {author} {\bibfnamefont {T.~N.}\ \bibnamefont
  {Bandi}}, \bibinfo {author} {\bibfnamefont {V.~G.}\ \bibnamefont {Minogin}},
  \ and\ \bibinfo {author} {\bibfnamefont {S.~N.}\ \bibnamefont {Chormaic}},\
  }\href@noop {} {\bibfield  {journal} {\bibinfo  {journal} {Physical Review A
  - Atomic, Molecular, and Optical Physics}\ }\textbf {\bibinfo {volume} {78}}
  (\bibinfo {year} {2008})}\BibitemShut {NoStop}%
\bibitem [{\citenamefont {Mu}\ \emph {et~al.}(2009)\citenamefont {Mu},
  \citenamefont {Lu}, \citenamefont {Xu}, \citenamefont {Ji.},\ and\
  \citenamefont {Yin}}]{Mu2009Generation}%
  \BibitemOpen
  \bibfield  {author} {\bibinfo {author} {\bibfnamefont {R.}~\bibnamefont
  {Mu}}, \bibinfo {author} {\bibfnamefont {J.}~\bibnamefont {Lu}}, \bibinfo
  {author} {\bibfnamefont {S.}~\bibnamefont {Xu}}, \bibinfo {author}
  {\bibfnamefont {X.}~\bibnamefont {Ji.}}, \ and\ \bibinfo {author}
  {\bibfnamefont {J.}~\bibnamefont {Yin}},\ }\href@noop {} {\bibfield
  {journal} {\bibinfo  {journal} {Journal of the Optical Society of America B}\
  }\textbf {\bibinfo {volume} {26}} (\bibinfo {year} {2009})}\BibitemShut
  {NoStop}%
\bibitem [{\citenamefont {Sondermann}\ \emph {et~al.}(2020)\citenamefont
  {Sondermann}, \citenamefont {Fischer},\ and\ \citenamefont
  {Leuchs}}]{Sondermann2020Prospects}%
  \BibitemOpen
  \bibfield  {author} {\bibinfo {author} {\bibfnamefont {M.}~\bibnamefont
  {Sondermann}}, \bibinfo {author} {\bibfnamefont {M.}~\bibnamefont {Fischer}},
  \ and\ \bibinfo {author} {\bibfnamefont {G.}~\bibnamefont {Leuchs}},\
  }\href@noop {} {\bibfield  {journal} {\bibinfo  {journal} {Advanced Quantum
  Technologies}\ }\textbf {\bibinfo {volume} {3}},\ \bibinfo {pages} {2000022}
  (\bibinfo {year} {2020})}\BibitemShut {NoStop}%
\bibitem [{\citenamefont {Vovrosh}\ \emph {et~al.}(2017)\citenamefont
  {Vovrosh}, \citenamefont {Rashid}, \citenamefont {Hempston}, \citenamefont
  {Bateman}, \citenamefont {Paternostro},\ and\ \citenamefont
  {Ulbricht}}]{Vovrosh2017Parametric}%
  \BibitemOpen
  \bibfield  {author} {\bibinfo {author} {\bibfnamefont {J.}~\bibnamefont
  {Vovrosh}}, \bibinfo {author} {\bibfnamefont {M.}~\bibnamefont {Rashid}},
  \bibinfo {author} {\bibfnamefont {D.}~\bibnamefont {Hempston}}, \bibinfo
  {author} {\bibfnamefont {J.}~\bibnamefont {Bateman}}, \bibinfo {author}
  {\bibfnamefont {M.}~\bibnamefont {Paternostro}}, \ and\ \bibinfo {author}
  {\bibfnamefont {H.}~\bibnamefont {Ulbricht}},\ }\href {\doibase
  10.1364/josab.34.001421} {\bibfield  {journal} {\bibinfo  {journal} {Journal
  of the Optical Society of America B}\ }\textbf {\bibinfo {volume} {34}},\
  \bibinfo {pages} {1421} (\bibinfo {year} {2017})}\BibitemShut {NoStop}%
\end{thebibliography}%


\end{document}